\documentclass[useAMS,usenatbib]{mn2e}

\usepackage{graphicx}

\newcommand{\change}{ }

\title[Spots on the surface of AR\,Aur]{Inhomogeneous surface distribution of chemical elements
in the eclipsing binary AR\,Aur:
A new challenge for our understanding of HgMn stars\thanks{Based on observations obtained at the European Southern
Observatory, Paranal, Chile (ESO programmes 268.D-5738(A), 076.D-0169(A), and 076.C-0170(A)).}}
\author[S. Hubrig et al.]{S. Hubrig$^{1}$\thanks{E-mail: shubrig@eso.org,},
 J.~F. Gonz\'alez$^{2}$, I. Savanov$^{3}$, M. Sch\"oller$^{1}$, N. Ageorges$^{1}$,\and
C.R. Cowley$^{4}$, and B. Wolff$^{5}$\\
$^{1}$European Southern Observatory, Casilla 19001, Santiago, Chile\\
$^{2}$Complejo Astron\'omico El Leoncito, Casilla 467, 5400 San Juan, Argentina\\ 
$^{3}$Astrophysical Institute Potsdam, An der Sternwarte 16, 14482 Potsdam, Germany\\
$^{4}$Department of Astronomy, University of Michigan, Ann Arbor, MI 48109-1042, USA\\
$^{5}$European Southern Observatory, Karl-Schwarzschild-Str. 2, 85748 Garching, Germany
}

\begin{document}

\date{Accepted 2006 Enero 99. Received 2006 Enero 98}

\pagerange{\pageref{firstpage}--\pageref{lastpage}} \pubyear{2006}

\maketitle

\label{firstpage}

\begin{abstract}
We present the results of a high spectral resolution study of the eclipsing binary AR\,Aur.
AR\,Aur is the only known eclipsing binary with a HgMn primary star exactly
on the ZAMS and a secondary star still contracting towards the ZAMS.
We detect for the first time  in the spectra of the primary star that for many elements 
the line profiles are variable over the rotation 
period. The strongest profile variations are found for the elements Pt, Hg, Sr, Y, Zr, He and Nd, while 
the line profiles of O, Na, Mg, Si, Ca, Ti, and Fe show only weak distortions over the rotation period.
The slight variability of  He and Y is also confirmed by the study of high resolution spectra
of another HgMn star: $\alpha$\,And. 
A preliminary modelling of the inhomogeneous distribution
has been carried out for Sr and Y.
Our analysis shows that these elements are very likely concentrated in a fractured ring along the 
rotational equator. 
It may be an essential clue for the explanation of 
the origin of the chemical anomalies in HgMn stars (which are very frequently found in binary and multiple 
systems) that one large fraction of the ring is missing exactly on the surface area which is permanently facing 
the secondary, and another small one on the almost opposite side.
The results presented about the inhomogeneous 
distribution  of various  chemical elements over the stellar surface of the primary suggest new 
directions for investigations to solve the question of the origin
of abundance anomalies in B-type stars with HgMn peculiarity.
\end{abstract}


\section{Introduction}

The origin of the abundance anomalies observed in late B-type stars with HgMn 
peculiarity is still poorly understood. 
Observationally, these stars are characterized
by low rotational velocities and weak or non-detectable magnetic fields.
Intrinsic photometric variability has been difficult to detect.
The most distinctive features of their atmospheres are an extreme
overabundance of Hg (up to 6 dex) and/or Mn (up to 3 dex) and a deficiency of He.
As more than 2/3 of the HgMn stars are known
to belong to spectroscopic binaries \citep{1995ComAp..18..167H},
the variation
of spectral lines observed in any HgMn star is usually explained to be due to
the orbital motion of the companion. 
The aspect of inhomogeneous distribution of some chemical 
elements over the surface of HgMn stars 
has been for the first time discussed by \citet{1995ComAp..18..167H}.
From the survey of HgMn stars in close SBs it was
suggested that some chemical elements might be inhomogeneously
distributed on the surface, with, in particular, preferential concentration
of Hg along the equator. In close SB2 
systems where the orbital plane has a small inclination to the line of sight,
a rather large overabundance of Hg 
was found. By contrast, in stars with
orbits almost perpendicular to the line of sight, mercury is not observed at all.
The first definitively identified spectrum
variability which is not caused by the companion has recently been 
reported for the binary HgMn star $\alpha$\,And by \citet{2001AAS...19913504W}
and \citet{2002ApJ...575..449A}.
They suggested that the spectral variations of the Hg~II line 
at $\lambda\,3984$\,\AA{} discovered in high-dispersion spectra are not due to
the orbital motion of the companion, but produced by the combination
of the 2.8\,day period of rotation of the primary and a nonuniform surface 
distribution of mercury which is concentrated in the equatorial region, in good
correspondence with the results of \citet{1995ComAp..18..167H}.

The ZAMS (Zero-Age Main Sequence) eclipsing binary AR\,Aur (HD\,34364, B9V+B9.5V) with an orbital 
period of 4.13\,days at an 
age of only $4\times{}10^6$\,years belongs to the Aur OB1 association \citep{1994A&A...282..787N}
and presents the best 
case for a study of evolutionary aspects of the chemical peculiarity phenomenon. This 
is the only eclipsing binary with HgMn peculiarity known to date.
The primary and the 
secondary eclipses are nearly total since its orbital inclination is 
88.5$^\circ$, and the radii of both stars are almost equal \citep{1994A&A...282..787N}.
\citet{1988BAICz..39...69C} discovered a third body in the system.
The existence of the as yet unseen 
third star with a mass of at least 0.51\,M$_\odot$ has been inferred from a light-time effect in the 
observed minima with a period of 25-27 years.
\citet{1994A&A...282..787N} studied the parameters of this multiple system in 
detail through an analysis of the available light and radial velocity
curves. They concluded that the secondary star is still contracting towards 
the ZAMS, while the primary star appears to be exactly on the ZAMS. 
Their data also
confirmed the existence of a third body with an orbital period of 24\,years and an 
orbital eccentricity of 0.17.
Remarkably, \citet{1979PASJ...31..821T} reported that AR\,Aur exhibits variations 
in the Hg\,II $\lambda\,3983.8$\,\AA{}  and Y\,II $\lambda\,3982.6$\,\AA{} line profiles. 
Just before the mid-secondary minimum, the Hg\,II line suddenly became the
strongest over all the observed phases, showing double-line structure. 
To explain the variations of mercury, Takada suggested that the primary star has some
inhomogeneities such as a cloud, a spot, or stratification.
Later, \citet{1997CoSka..27...41Z} reached a similar
conclusion about the variable profiles of mercury. However, both studies used
photographic observations at a low spectral resolution of about 30,000 or less and a S/N of $\sim$50.
Certainly, a more rigorous study using high quality, high-resolution spectroscopy at a very high 
S/N ratio should provide convincing evidence that the Hg\,II and Y\,II line profiles are variable.
Therefore, we decided to obtain high quality UVES spectra of this system to allow
a careful study of the previously reported anomalous line profile characteristics of these 
elements. 

\section{Observations}

Nine spectra of  AR\,Aur have been
recorded 
at ESO with the VLT UV-Visual Echelle Spectrograph
UVES at UT2 in 2005. We used the UVES dichroic standard settings covering the spectral 
range from 3030\,\AA{} to 10,000\,\AA{}. 
The slit width was set to $0\farcs{}3$
for the red arm, corresponding to a resolving power of 
$\lambda{}/\Delta{}\lambda{} \approx 1.1\times10^5$. For the blue arm, we used 
a slit width of $0\farcs{}4$ to achieve a resolving power of 
$\approx 0.8\times10^5$.
The spectra have been reduced by the UVES pipeline Data Reduction Software
(version 2.5; \citealt{2000Mess.101..31}) and using IRAF.
The signal-to-noise ratios of the resulting UVES spectra range
from 210 to 380 per pixel in the one-dimensional spectrum around 
3990\,\AA{}, {\change with the maximum S/N of about 450 around 4500\,\AA{}. }
The summary of our spectroscopic observations of AR\,Aur is given in Table\,1.

The theoretical value for synchronous rotation of AR\,Aur is about 22\,km\,s$^{-1}$ \citep{1994A&A...282..787N}. 
The line widths in the spectra of both components are nearly identical. Our 
estimates using eleven Fe\,II and Ti\,II lines in the region $\lambda\lambda\,5000-5300$\,\AA{} result 
in $v \sin i = 22\pm{}1$\,km\,s$^{-1}$. This 
value is consistent with the assumption of synchronous rotation.
Therefore, the rotation phases have been calculated assuming $P_{\rm rot}=P_{\rm orb}=4.13$\,days.
The ephemeris have been taken from the most recent study of AR\,Aur by  \citet{2003AN....324..523H}.
Unfortunately, since our observations have been carried out randomly in service mode, no spectrum 
was obtained at the phases of eclipses. Furthermore, three spectra have been taken around
rotation phase 0.1 and two spectra around phase 0.8. Hence, we 
have only a coverage over six different rotation phases, which is not sufficient to 
perform a reconstruction of the surface abundance distribution from the rotationally 
modulated spectral line profiles using the Doppler 
Imaging technique. On the other hand, the observations we have obtained so far allow us to detect 
spectral line profile variations, if present, and to carry out modelling 
of spectral line profiles from test images with certain abundance distributions. 


\begin{table}
\label{tab:results} \caption{The log of the observations of AR\,Aur with UVES. The signal-to-noise 
ratio refers to the spectral region around 3990\,\AA{}.}
\begin{center}
\begin{tabular}{ccc|ccc}
\hline
\multicolumn{1}{c}{JD}  &
\multicolumn{1}{c}{S/N} &
\multicolumn{1}{c}{Phase} &
\multicolumn{1}{c}{JD}  &
\multicolumn{1}{c}{S/N} &
\multicolumn{1}{c}{Phase} \\
\multicolumn{1}{c}{- 2,450,000}  & & & 
\multicolumn{1}{c}{- 2,450,000}  & & \\
\hline
3632.871& 380& 0.658 & 3724.613&210& 0.846\\
3671.821&305& 0.078 & 3725.621&360& 0.091\\
3700.768&200& 0.079 & 3728.587&240& 0.807\\
3713.653&270& 0.196 & 3730.589&250& 0.292\\
3722.649&275& 0.371\\
\hline
\end{tabular}
\end{center}
\end{table}

\begin{figure*}
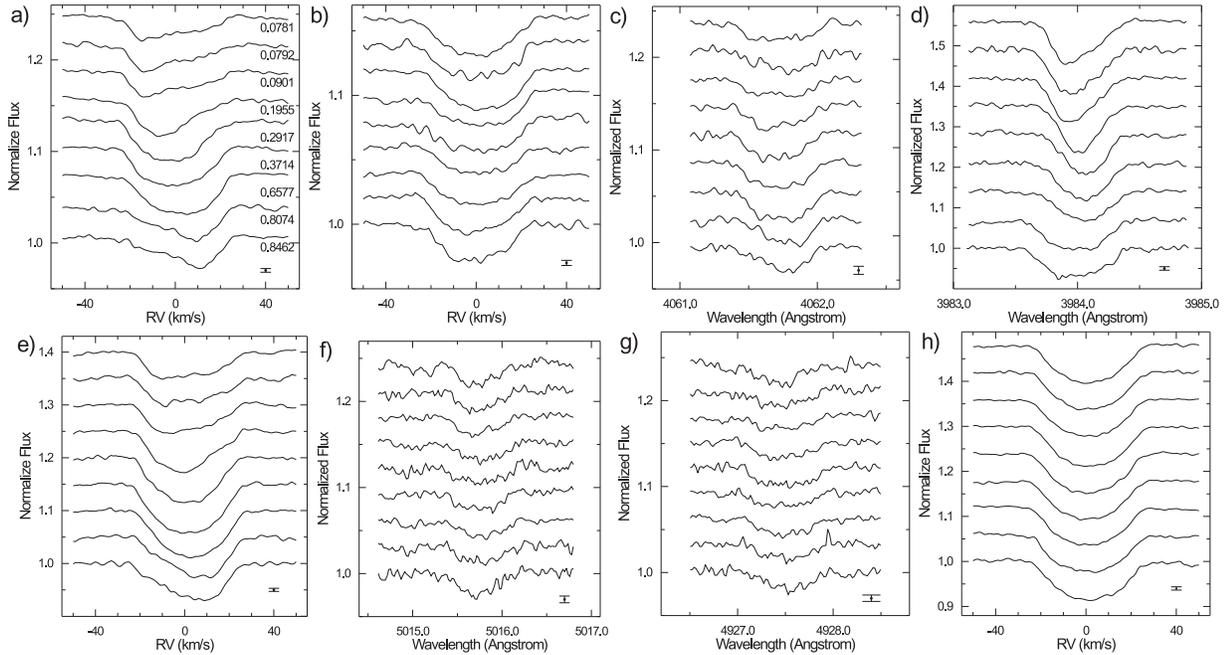

\centering
\includegraphics[width=0.22\textwidth]{fig1a-Yphase.eps}
\includegraphics[width=0.22\textwidth]{fig1b-Zr.eps}
\includegraphics[width=0.22\textwidth]{fig1c-Pt.1.eps}
\includegraphics[width=0.233\textwidth]{fig1d-Hg.eps}
\includegraphics[width=0.22\textwidth]{fig1e-Sr.eps}
\includegraphics[width=0.22\textwidth]{fig1f-He.eps}
\includegraphics[width=0.22\textwidth]{fig1g-Nd.eps}
\includegraphics[width=0.22\textwidth]{fig1h-Fe.eps}
\caption{Variations of line profiles phased with the rotation period P = 4.13\,days:
a)~Y\,II,
b)~Zr\,II,
c)~Pt\,II $\lambda\,4061.7$,
d)~Hg\,II $\lambda\,3983.9$,
e)~Sr\,II,
f)~He\,I $\lambda\,5015.7$,
g)~Nd\,III $\lambda\,4927.5$, and
h)~Fe\,II.
The rotational phase increases from top to bottom - see  Fig.\,\ref{fig1}a. Error bars in the 
lower right corner indicate the standard error of 
the line profiles. Individual spectra are shifted in vertical direction.}
\label{fig1}  
\end{figure*}

\section{Spectral variability and modelling}

The complete set of abundances in the atmospheres of both primary and secondary components 
of the AR\,Aur system can be found in the paper of \citet{1995AstL...21..818K}.
Of special interest is the fact that lines of Y, Zr, Nd, Pt, and Hg are only present in the spectra of
the primary component. Remarkably, inspecting the behavior of spectral 
lines at different rotation phases we discovered that exactly these elements 
show a distinct spectral variability. The lines of a few other elements (O, Na, Mg, Si, Ca, Ti, Fe, 
Sr, and He) 
are present in the spectra of the primary and the secondary component, but they appear variable only in the 
spectra of the primary. No noticeable variability of any element has been detected in the spectra of 
the secondary star.

To study the behavior of the line profiles
and to get a better idea about the distribution of various elements on the surface of the primary star,
we decided to compute for a few elements, namely Y, Zr, Sr, and Fe,  mean line profiles,
where a set of unblended lines of similar strength is available.
We note that mean 
profiles can not be used for Hg and Pt due to different isotopic and hyperfine structures of 
individual Hg and Pt lines.
We also checked whether the discovered variability of spectral lines is not caused by 
mutual blending of lines originating in the individual binary components. 

The problem of analyzing the component spectra in double-lined spectroscopic binaries is a 
tough one, but, fortunately, in the past few years several techniques for spectral disentangling 
have been developed. We applied the procedure of decomposition described in detail by 
\citet{2006A&A...448..283G} for each observed phase and could fully confirm the detected 
variability of the lines of all elements mentioned above.
In Fig.\,\ref{fig1} we show the behavior of the line profiles of some elements over the rotation period.
\begin{figure}
\centering
\includegraphics[width=0.43\textwidth]{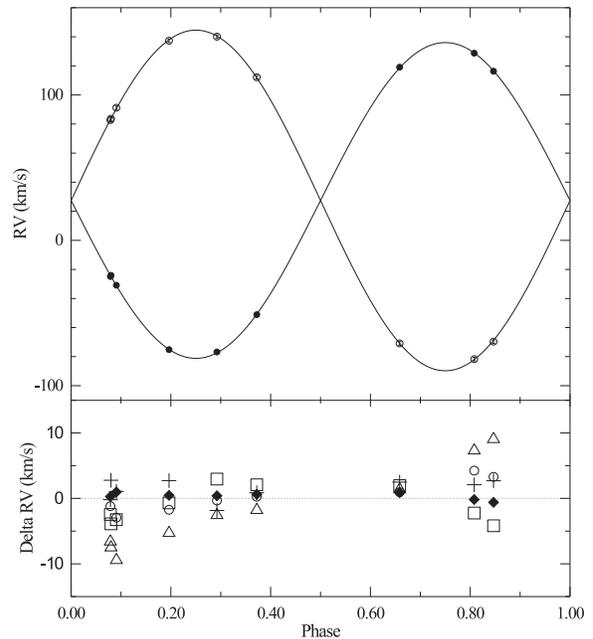}
\caption{
Radial velocity curves of AR\,Aur. Small filled and
open circles are the mean
stellar radial velocity of the primary and secondary components,
respectively. In the lower panel we show the radial velocity difference of
various elements with
respect to the mean stellar velocity of the primary. Squares present 
the measurements for the Hg\,II line $\lambda\,3984$, triangles are the results for Y\,II,
open circles for Sr\,II, pluses for Ca\,II, and filled diamonds for Ti\,II lines.
The rms of the residuals of the radial velocity curves is about 0.5\,km\,s$^{-1}$. The errors in the
radial velocities of individual elements are of the order of 1\,km\,s$^{-1}$ for slightly variable elements and 
about 2\,km\,s$^{-1}$ for strongly variable elements.
}
\label{fig2}  
\end{figure}
The Zr\,II, Nd\,III, Pt\,II, and He\,I lines appear rather weak, but still their variations are 
well noticeable.
Interestingly, while the behavior of the line profiles of Y\,II, Pt\,II,
Hg\,II, Sr\,II, and Nd\,III is rather similar over the rotation period, the line profiles of Zr\,II and He\,I
seem to vary with a 180$^\circ$ phase shift.
In Fig.\,\ref{fig1}h we display the mean line profiles of Fe\,II obtained using six unblended lines.
Although we see very weak distortions in the individual lines over the rotation period,
the width of the mean line profile suggests that Fe is distributed almost 
homogeneously over the stellar surface. The same kind of slight distortions is also 
observed in the O\,I, Na\,I, Mg\,II, Si\,II, Ca\,II, and Ti\,II line profiles.
Moreover, the comparison of the observed line profiles
with  synthetic spectra calculated assuming the atmosphere parameters employed by \citet{1994A&A...282..787N}
($T_{\rm eff}=10950$\,K, $\log g=4.33$)
and assuming $v_{\rm t}=0.5$\,km\,s$^{-1}$) 
reveals that for these elements the overall
shape of  the line profiles slightly deviates from the purely rotationally broadened profiles. 
In the spectrum 
recorded at rotational phase 0.85, the Hg\,II line at $\lambda\,3984$
appears very broad (Fig.\,\ref{fig1}d), even broader than the mean line profiles of Fe\,II. 
As the separation of the 
isotope components of this Hg\,II line is of the same order as the Doppler broadening, such a 
shape of 
the line profile could be interpreted as a signature of the unusual isotopic mixture in that 
rotation phase.
However, this explanation has to be proven by a detailed analysis of the Hg isotopic composition and
application of the Doppler Imaging reconstruction method.

In Fig.\,\ref{fig2} we show radial velocity curves of AR\,Aur. For a few elements 
which are distributed inhomogeneously over the stellar surface of the primary, we present in 
the lower panel 
the radial velocity differences (elements Hg, Y, Sr, Ca, and Ti). Since most of the line profiles 
are asymmetric, the center-of-gravity method has been used 
to measure the precise line position. However, the measurements are less reliable for 
the Hg\,II line $\lambda\,3984$, as discussed above.

\begin{figure}
\centering
\includegraphics[width=0.22\textwidth]{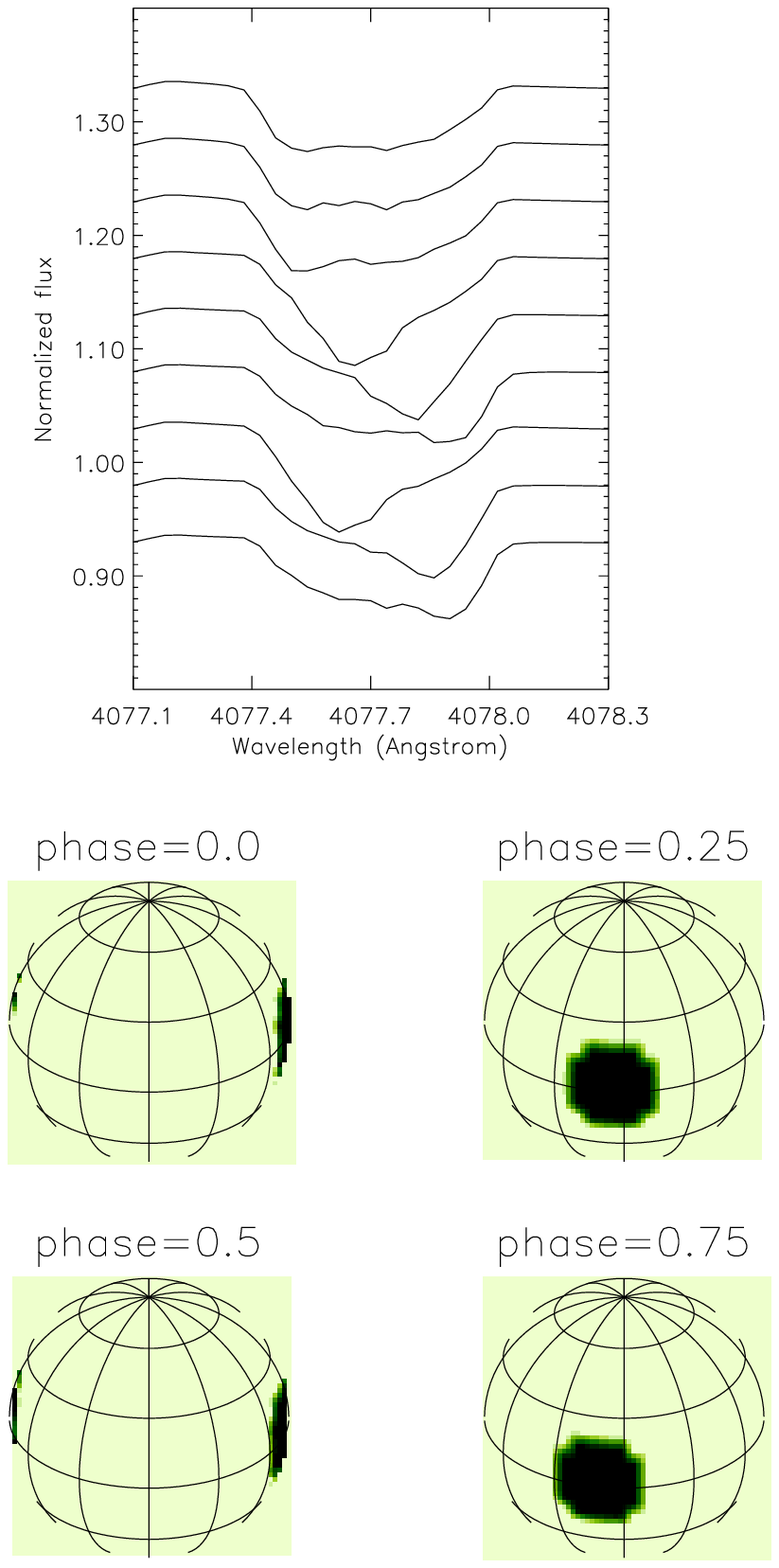}
\includegraphics[width=0.22\textwidth]{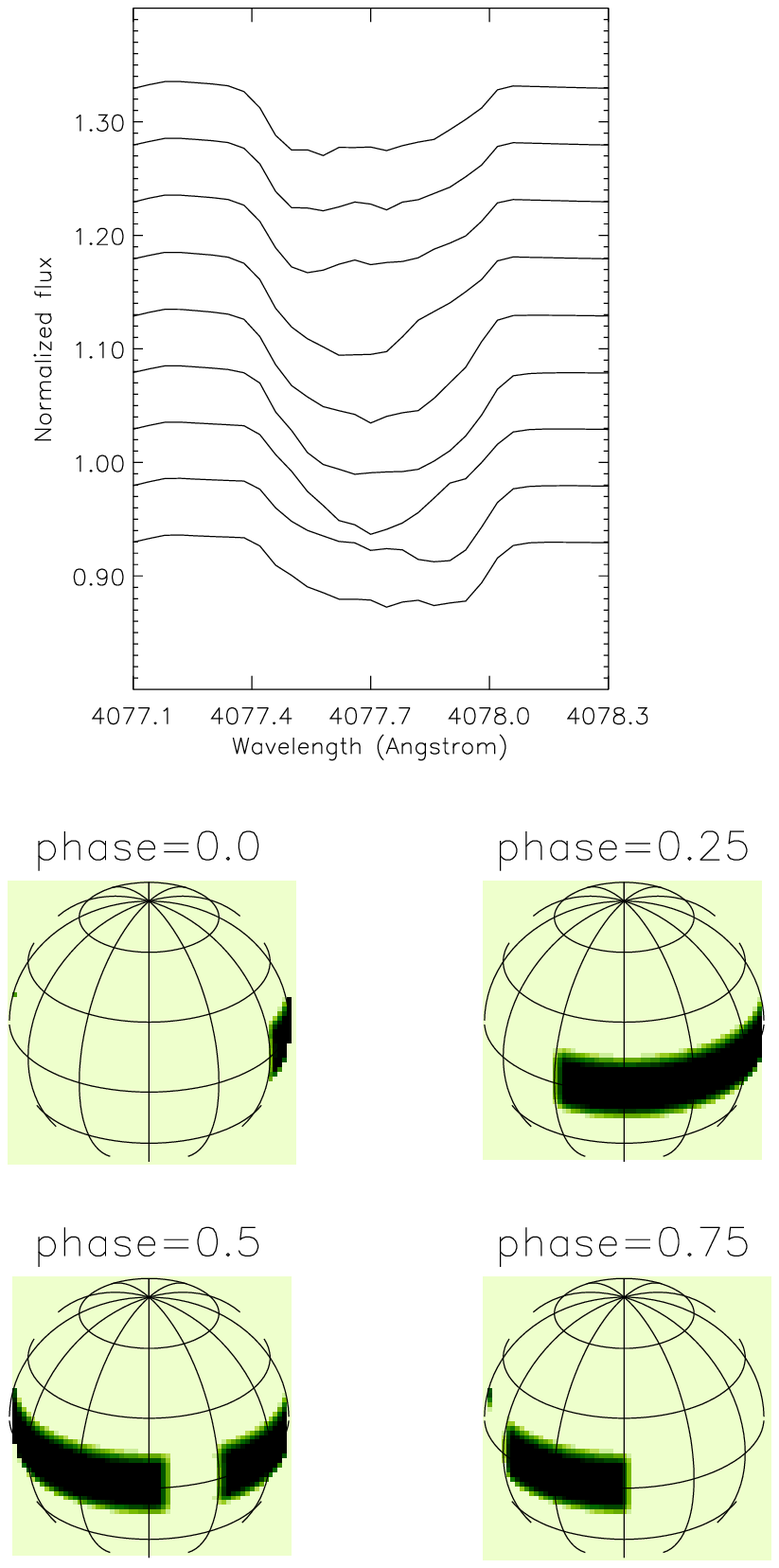}
\caption{
Sr synthetic spectra (top) and Sr maps (bottom) for the two models considered:
two spots on the surface (left) and a fractured ring (right).
}
\label{fig3}  
\end{figure}

In the following we present the first results of our 
preliminary modelling of abundance distributions of Sr\,II and Y\,II using 
the Sr\,II line $\lambda\,4078$ and the Y\,II line $\lambda\,4900$.
We used the so-called direct Doppler Imaging 
method recently presented by \citet{2006A&A...444..931x}.
This method allows to compute spectral line profiles from test images 
created by varying the local element abundances (or surface temperature 
inhomogeneities) and their location on the stellar surface.    
The model atmosphere has been  
calculated with Kurucz's ATLAS9 program \citep{1993KurCD..13.....K}
and the local line profiles have been synthesized with the 
program SYNTH
\citep{1992stma.conf...92P}.
The VALD database (\citealt{1999A&AS..138..119K})
provided the atomic line data necessary for the spectral synthesis 
of the local line profiles. The results of modelling two almost 
opposite equatorial spots with radius R$\sim$18$^\circ$ located at stellar longitudes 
$l_{\rm 1}$$\sim$84$^\circ$ and $l_{\rm 2}$$\sim$258$^\circ$, 
are presented for Sr in Fig.\,\ref{fig3} on the left.
Sr was found overabundant in the atmosphere by 0.5\,dex 
and by about 3.0\,dex in spots, compared to the solar value.
For Y we found an overabundance in the atmosphere by 2.0\,dex 
and by about 5.0\,dex in the spots.
However, an employment of a partially 
fractured equatorial ring with a width of $\sim$24$^\circ$ (Fig.\,\ref{fig3}, right)
presents a much better consistency between the calculated line profiles and the observed ones. 
Therefore we conclude that the ring structure presents a more realistic 
distribution  of Sr and Y on the stellar surface.
It may be an essential clue for the understanding of 
the origin of the chemical anomalies in HgMn stars (which are very frequently found in binary and multiple 
systems) that one large fraction of the ring is missing exactly on the surface area which is permanently facing 
the secondary, and another small one on the almost opposite side. It is intriguing that the results of 
Hg mapping on the surface of $\alpha$\,And presented 
by \citet{2001AAS...19913504W} and \citet{2002ApJ...575..449A}
are rather similar, showing a concentration of Hg in a somewhat non-uniform equatorial band.
From evaluating the changes of the line profiles over the rotation period in the spectra of AR\,Aur, it 
seems that the distribution of Hg, Pt, and Nd shows a behavior comparable to Sr and Y.

\begin{figure*}
\centering
\includegraphics[width=0.8\textwidth]{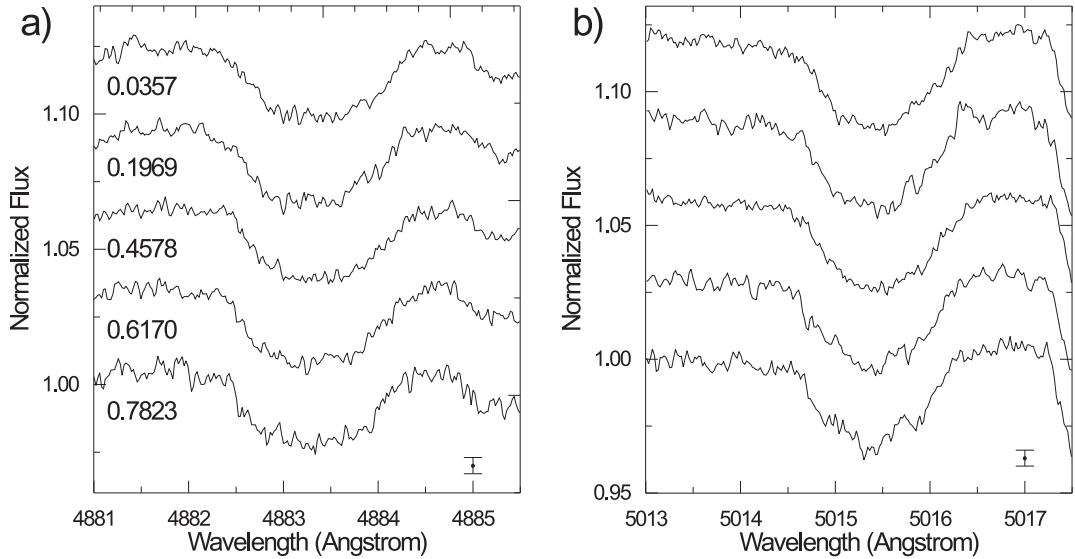}
\caption{
Spectral variability of $\alpha$\,And: a)~Variability of Y\,II $\lambda\,4883.7$, 
b)~Variability of He\,I $\lambda\,5015.7$.
{\change Error bars in the lower right corner indicate the standard error of 
the line profiles.}}
\label{fig4}  
\end{figure*}

Since a few years we have at our disposal 
a few spectra of the HgMn star $\alpha$\,And, recorded
during five consecutive nights in 2001. Similar to the observations 
of AR\,Aur we used the UVES dichroic standard settings covering the spectral 
range from 3030\,\AA{} to 10,000\,\AA{}, giving a S/N {\change greater} than 400 in the spectral region 
around 3990\,\AA{} and more than 500 around 4500\,\AA{}.
From the inspection of these spectra
we see the variability of Hg previously announced
by \citet{2001AAS...19913504W} and \citet{2002ApJ...575..449A}.
In addition, we find that also He\,I and Y\,II lines 
appear slightly variable. If these elements are concentrated in a fractured equatorial ring, similar to that 
in AR\,Aur,  the significantly 
lower variability of Y and He in $\alpha$\,And 
could be explained by a different inclination (74$^\circ$) of the rotational axis to the line of sight.
The variability of the line profiles $\lambda\,5015.6$ He\,I and $\lambda\,4884$ Y\,II 
in the spectra of $\alpha$\,And is presented in Fig.\,\ref{fig4}. 

The primary and the secondary components in the AR\,Aur system  are  of considerable interest as they are very 
young and nearly equal-mass stars. 
\citet{1994A&A...282..787N} suggested that the third star
could be a pre-main sequence star.
A direct detection of the third component would support this scenario 
and provide an estimate of the mass of this component and the inclination between the two orbital planes.
We obtained high-resolution images of AR\,Aur in November 2005 with NACO at the VLT.
The observations were carried out in the $K_{\rm s}$ band using the camera S13. 
However, no companion has been detected.
We note that the orbital inclination of the wide 
orbit is unknown.
As we are presently almost in the phase of conjunction 
the companion would be within the diffraction limit for inclinations below
50$^\circ$.
Additionally, the third component could be too faint to detect in close proximity
to the two more massive stars.
\section{Discussion}

The discovery of an inhomogeneous distribution of various elements in the atmospheres of AR\,Aur and
$\alpha$\,And challenges our understanding of the nature of HgMn stars. 
Using FEROS at the ESO 2.2\,m telescope we are currently carrying out a survey of a large sample of HgMn 
stars to establish the spectral variability due to an inhomogeneous elemental
distribution in the atmospheres of these stars. The observations obtained during the last 
months  led already to the
discovery of a number of HgMn stars with spots (Gonz\'alez et al. in preparation), proving
that the presence of an inhomogeneous distribution on the surface of these stars is a rather common 
characteristics and not just a rare phenomenon found in the atmospheres of very few HgMn stars.
The results of our study modify the claim of \citet{2002ApJ...575..449A}
that the variability 
phenomenon in HgMn stars is confined to a narrow range of stellar parameters rather than 
characterizing the entire class of HgMn stars.

Presently we do not have any explanation for the discovered distribution of Sr and Y in a fractured ring 
where a large fraction of the ring is missing exactly on the surface area which is permanently facing 
the secondary, and another small fraction in the almost opposite phase.
We believe that factors as the presence of a weak tangled magnetic field, tidal distortion, or 
the reflection effect can play a role in the development of anomalies in HgMn stars. 
Although diffusion due to gravitational settling and radiative levitation have been 
the most popular explanation for the HgMn star abundances in the last decades, the accretion 
of interstellar material during the pre-main-sequence phase by HgMn binary systems seems to be a 
more convincing choice to explain the surface abundances in the presence of magnetic fields.
Probably, all these 
mechanisms have to be taken into account in future studies of these stars.

Typically, inhomogeneous chemical abundance distributions are observed only on the surface of magnetic 
chemically peculiar stars with large-scale organized magnetic fields. In these stars
the abundance distribution of certain elements is non-uniform and non-symmetric with respect to 
the rotation axis.
The presence of  
magnetic fields in the atmospheres of HgMn stars has been studied by
\citet{1995A&A...293..810M} and \citet{2001A&A...375..963H}.
\citet{1995A&A...293..810M} could demonstrate the presence of quadratic magnetic fields 
in two close double-lined systems with HgMn primary stars, 74\,Aqr and $\chi$\,Lup.
\citet{2001A&A...375..963H} showed for a few HgMn stars
evidence for a relative magnetic intensification of Fe\,II lines
produced by different magnetic desaturations induced by different
Zeeman-split components.
As the relative intensification is roughly correlated
with the strength of the magnetic field, it is a powerful tool for detecting
magnetic fields which have a complex structure and are difficult to detect by
polarization measurements.
A magnetic field of the order of a few hundred Gauss was recently detected in four other HgMn stars
by \citet{2006AN....327..289H}.
The small sample of HgMn stars for which \citet{2006AN....327..289H} gathered 
magnetic field measurements with FORS\,1 in spectropolarimetric 
mode also included $\alpha$\,And, with a detected  variable magnetic field of the order 
of up to 260\,G. 
The structure of the measured field in HgMn stars must be, however,
sufficiently tangled so that it does not produce a strong net observable
circular polarization signature. A scenario how a magnetic field can be build up in binary 
systems has been presented some time ago by \citet{1998CoSka..27..296H}.
As the role that magnetic fields play in the development 
of HgMn chemical peculiarities is unclear yet, systematic searches for magnetic fields in 
a larger sample of these stars should be conducted and the structure of those magnetic fields 
needs to be determined.

Finally, we would like to note that since
many stars with HgMn peculiarity are found in young multiple systems and in young associations, 
further studies of HgMn stars are especially important 
to understand the evolutionary aspects of the chemical peculiarity phenomenon.
Our recent study of the multiplicity of HgMn stars suggests that at least 2/3 of HgMn stars are formed 
in multiple systems \citep{2006msahrd...}.
It is conceivable that HgMn stars have been synchronized and 
that their peculiarities were established as early as the PMS phase. Both AR\,Aur and $\alpha$\,And appear 
very young, with the secondary in the AR\,Aur system in the final stage of contracting towards 
the ZAMS. $\alpha$\,And was also found to be located on the ZAMS \citep{2005ASPC..337..236H}.


\label{lastpage}


\begin{thebibliography}{99}

\bibitem[\protect\citeauthoryear{Adelman et al.}{2002}]{2002ApJ...575..449A}
Adelman S.~J., Gulliver A.~F., Kochukhov O.~P., Ryabchikova T.~A., 2002, ApJ, 575, 449 

\bibitem[\protect\citeauthoryear{Albayrak et al.}{2003}]{2003AN....324..523H} 
Albayrak B., Ak T., Elmasli A., 2003, AN, 324, 523

\bibitem[\protect\citeauthoryear{Ballester et al.}{2000}]{2000Mess.101..31}
Ballester P., Modigliani A., Boitquin O., Cristiani S., et al., 2000, Messenger, 101, 31

\bibitem[\protect\citeauthoryear{Chochol et al.}{1988}]{1988BAICz..39...69C}
Chochol D., Juza K., Zverko J., Ziznovsky J., Mayer P., 1988, BAICz, 39, 69 

\bibitem[\protect\citeauthoryear{Gonz{\'a}lez \& Levato}{2006}]{2006A&A...448..283G}
Gonz{\'a}lez J.~F., Levato H., 2006, A\&A, 448, 283 
\bibitem[\protect\citeauthoryear{Hubrig}{1998}]{1998CoSka..27..296H}
Hubrig S., 1998, CoSka, 27, 296 

\bibitem[\protect\citeauthoryear{Hubrig et al.}{2006a}]{2006msahrd...} 
Hubrig S., Ageorges N., Sch{\"o}ller M, 2006a,
in 
ESO Astrophys. Symposia,
Multiple Stars across the H-R diagram.
Springer, Berlin/Heidelberg, in press

\bibitem[\protect\citeauthoryear{Hubrig \& Castelli}{2001}]{2001A&A...375..963H}
Hubrig S., Castelli F., 2001, A\&A, 375, 963 

\bibitem[\protect\citeauthoryear{Hubrig \& Mathys}{1995}]{1995ComAp..18..167H}
Hubrig S., Mathys G., 1995, Com. Ap, 18, 167 

\bibitem[\protect\citeauthoryear{Hubrig et al.}{2006b}]{2006AN....327..289H} 
Hubrig S., North P., Sch{\"o}ller M., Mathys G., 2006b, AN, 327, 289 

\bibitem[\protect\citeauthoryear{Hubrig et al.}{2005}]{2005ASPC..337..236H} 
Hubrig S., Szeifert T., North P., Sch{\"o}ller M., Yudin R.~V., 2005, ASPC, 337, 236 

\bibitem[\protect\citeauthoryear{Khokhlova et al.}{1995}]{1995AstL...21..818K}
Khokhlova V.~L., Zverko Y., Zhizhnovskii I., Griffin R.~E.~M., 1995, AstL, 21, 818 

\bibitem[\protect\citeauthoryear{Kupka et al.}{1999}]{1999A&AS..138..119K} 
Kupka F., Piskunov N., Ryabchikova T.~A., Stempels H.~C., Weiss W.~W., 1999, A\&AS, 138, 119 

\bibitem[\protect\citeauthoryear{Kurucz}{1993}]{1993KurCD..13.....K}
Kurucz R., 1993, KurCD, 13, 

\bibitem[\protect\citeauthoryear{Mathys \& Hubrig}{1995}]{1995A&A...293..810M}
Mathys G., Hubrig S., 1995, A\&A, 293, 810 

\bibitem[\protect\citeauthoryear{Nordstrom \& Johansen}{1994}]{1994A&A...282..787N}
Nordstrom B., Johansen K.~T., 1994, A\&A, 282, 787 

\bibitem[\protect\citeauthoryear{Piskunov}{1992}]{1992stma.conf...92P} 
Piskunov N.~E., 1992,
in Glagolevskij Yu.V., Romanyuk I.I., eds,
Stellar Magnetism. NAUKA, Sankt-Petersburg, p.\ 92
\bibitem[\protect\citeauthoryear{Savanov \& Strassmeier}{2006}]{2006A&A...444..931x}
Savanov I., Strassmeier K., 2006, A\&A, 444, 931

\bibitem[\protect\citeauthoryear{Takeda et al.}{1979}]{1979PASJ...31..821T}
Takeda Y., Takada M., Kitamura M., 1979, PASJ, 31, 821

\bibitem[\protect\citeauthoryear{Wahlgren et al.}{2001}]{2001AAS...19913504W}
Wahlgren G.~M., Ilyin I., Kochukhov O., 2001, AAS, 33, 1506

\bibitem[\protect\citeauthoryear{Zverko et al.}{1997}]{1997CoSka..27...41Z}
Zverko J., Ziznovsky J., Khokhlova V.~L., 1997, CoSka, 27, 41

\end{thebibliography}
\end{document}